\documentclass[useAMS,usenatbib,usegraphicx]{mn2e}

\usepackage{times}
\usepackage[utf8x]{inputenc}

\newcommand{\fe}{Fe K$\alpha$ }
\newcommand{\etal}{et al.}

\title[The effects of dust on the equivalent width of the \fe\ line in
AGNs]
  {On the equivalent width of the \fe\ line produced by a dusty
    absorber in active galactic nuclei}
\author[R.\ Gohil and D.\ R.\ Ballantyne]
  {R.~Gohil\thanks{raj.gohil07@gmail.com} and D.~R.~Ballantyne\\Center
    for Relativistic Astrophysics, School of Physics, Georgia
    Institute of Technology, 837 State Street, Atlanta, GA 30332-0430,
    USA}

\pagerange{\pageref{firstpage}--\pageref{lastpage}}
\pubyear{2014}

\begin{document}

\label{firstpage}

\maketitle

\begin{abstract}

Obscured AGNs provide an opportunity to study the material surrounding
the central engine. Geometric and physical constraints on the absorber
can be deduced from the reprocessed AGN emission. In particular, the
obscuring gas may reprocess the nuclear X-ray emission producing a
narrow \fe line and a Compton reflection hump. In recent years, models
of the X-ray reflection from an obscuring torus have been computed;
however, although the reflecting gas may be dusty, the models do not
yet take into account the effects of dust on the predicted
spectrum. We study this problem by analyzing two sets of models, with
and without the presence of dust, using the one dimensional
photo-ionization code Cloudy. The calculations are performed for a
range of column densities ($22 <{\rm log}[N_H(\rm cm^{-2})]< 24.5$ )
and hydrogen densities ( $6 <{\rm log}[n_H(\rm cm^{-3})]< 8$). The
calculations show the presence of dust can enhance the \fe equivalent
width (EW) in the reflected spectrum by factors up to $\approx$ 8 for
Compton thick (CT) gas and a typical ISM grain size distribution. The
enhancement in EW with respect to the reflection continuum is due to
the reduction in the reflected continuum intensity caused by the
anisotropic scattering behaviour of dust grains. This effect
will be most relevant for reflection from distant,
predominately neutral gas, and is a possible explanation for AGNs which show a strong \fe EW and a relatively weak reflection continuum. Our results show it is an important to take into account dust while modeling the X-ray reflection spectrum, and that inferring a CT column density from an observed \fe EW may not always be valid. Multi-dimensional models are needed to fully explore the magnitude of the effect.

\end{abstract}

\begin{keywords}
galaxies:active-galaxies:Seyfert-X-rays
\vspace{0.2cm}
\end{keywords}

\section{Introduction}
\label{sec:intro}

All active galactic nuclei (AGN) are powered by gas accreting onto a central supermassive black hole \citep[e.g.][]{balbus03}, a process which emits a significant amount of energy across the electromagnetic spectrum. Interestingly, a significant number of AGNs show the presence of local obscuration \citep{comastri04} at distances $\sim$ 1-10 pc from the central engine \citep[e.g.][]{antonucci93, urry95}. The nature and origin of the absorbing gas is largely unknown but the gas presents an unique opportunity to study material as it transitions from the galaxy to the AGN environment.

Superimposed onto the typical $\Gamma$ $\sim$ 1.8-2 X-ray power-law in AGNs \citep[e.g.][]{dadina08, beckmann09, corral11}
is an \fe line \citep[e.g.][]{nandra94, ebisawa96}. The narrow
component of the line is observed in a large number of objects
\citep{yaqoob01, kaspi01}, even in many high-redshift samples
\citep{brusa05, corral08, chaudhary12, iwasawa12}, and the width of
the line indicates that it may originate in the obscuring material
\citep[e.g.][]{yaqoob04, shu10}. Moreover, the equivalent width (EW)
of the \fe line can be as large as several keV
\citep[e.g.][]{levenson06} which may be a signature of Compton thick
(CT) gas \citep[e.g.][]{murphy09}. A correlation between the \fe
  EW and the line-of-sight column density is observed for columns
  $>10^{23}$~cm\textsuperscript{-2} indicating that the line is often
  produced by CT gas in the AGN environment
  \citep[e.g.][]{guainazzi05,fukazawa10}. Thus, the \fe line can be
used as an important proxy to study the properties of the obscuring
gas around AGNs.

In recent years, there have been several Monte-Carlo models of X-ray reprocessing from a CT torus \citep[e.g.][]{ikeda09, murphy09, brightman11a}. However, all of these calculations omit dust from the models which in many cases may be present in the X-ray absorbing gas \citep{jaffe04,prieto04,prieto05,meisenheimer07,tristram07,raban09}. As these models are now being used to determine the gross properties of the obscuring material, it is important to consider the effects of dust on the X-ray reflection spectrum. Here, we use Cloudy models to show that the inclusion of dust in the distant obscuring gas may have an important impact on the reflected \fe EW. Sect. 2 describes the geometrical set up of the models and the calculation procedure. Sect. 3 describes the results, and Sect. 4 discusses the implications of this study.

\section{Calculations}
\label{sec:calc}

The photo-ionization code Cloudy v13.02 \citep{ferland13}, is utilized to study the impact of dust on the reflected iron K$\alpha$ line. Although the calculations with Cloudy are one dimensional, the code includes multiple grain species with a realistic size distribution \citep[e.g.][]{mathis77} as well as a comprehensive treatment of grain physics \citep{draine01b,vanhoof04,weingartner06}, i.e. scattering and absorption of dust based on its size and shape. All calculations assume an AGN spectrum with an intrinsic luminosity of 10\textsuperscript{43} erg s\textsuperscript{-1} in the 2-10 keV band and a typical photon index of 1.9 \citep[e.g.][]{beckmann09}. The optical to UV spectrum is characterized by  ${\alpha}_{ox}$ = -1.4 \citep[e.g.][]{zamorani81}, a UV slope of -0.5 \citep[e.g.][]{francis93,elvis94}, and a blackbody temperature of 1.4$\times10^5$ K (appropriate for a black hole of 10\textsuperscript{7} $M_\odot$ accreting at 10\% of its Eddington rate). The gas irradiated by this spectrum has an uniform hydrogen density, $n_H$, extended radially outward until it reaches a specified column density, $N_H$. The covering factor is fixed at 0.67, equivalent to a 30$^{\circ}$ opening angle, and calculations are performed for $22 <{\rm log}[N_H(\rm cm^{-2})]< 24.5$ (with 0.25 dex spacing), and $6 <{\rm log}[n_H(\rm cm^{-3})]< 8$ (with a spacing of 1 dex).

To study the effects of dust, two sets of models are computed. The first has gas with solar abundances \citep{grevesse10} and no dust (ND), while the second includes ISM dust (WD) \citep{mathis77,cowie86,savage96,meyer98,snow07}. In order to isolate the effects of dust on the reflected spectra, the abundances in the gas in the WD models are kept at solar, but the metal abundances are reduced using the depletion factors of \cite{cowie86}, and \cite{jenkins87} and the gas-to-dust ratio was kept at 151. To check the self consistency of the two models, the abundances of each element are added and the overall abundance of elements is consistent within 0.05\% in both ND and WD models; however, the Fe abundance in the WD models is 5\% higher than ND models.

At high AGN luminosities grains may reach the sublimation temperature and return the metals back to the gas phase. This effect is not easily treated by Cloudy, so the inner radius of the illuminated gas is fixed at a distance of 10 pc away from the AGN. At this distance the dust temperature inside the illuminated gas is always below the sublimation temperature of the grains. As a further check on the choice of parameters, the reprocessed 12$\mu$m luminosity of the WD models is compared against the one expected from the 12 $\rm{\mu}$m-X-ray luminosity relationship \citep{gandhi09}. Depending on $n_H$, the simulated 12$\mu$m luminosities are 1.2-1.5 times larger than the observationally predicted one. Thus, our simple dusty gas setup will be a reasonable model for the average properties of a Seyfert-type AGN.

Finally, to study the effects of dust on the Fe K$\alpha$ emission line, the equivalent width (EW) of the line is analyzed as a function of $N_H$ and $n_H$ for both sets of models. The EW is defined as the ratio of the intensity of the Fe K$\alpha$ emission line ($I_{Fe K\alpha}$) to the reflected continuum at the line energy ($I_c$) multiplied by the energy bin width ($\Delta E$):

\begin{equation}
\label{equ2}
\centering
EW= \frac{I_{Fe K\alpha}}{I_c}\times\Delta E. 
\end{equation}
Ideally, $I_c$ should be measured at 6.4 keV, but because of the presence of the emission line, it is difficult to determine $I_c$ at that energy. Therefore, $I_c$ is measured at 6.3 keV as there are no predicted emission lines at that energy. Using a different $I_c$ between 6.3  keV and 6.5 keV leads to only very small changes in the EW. Rather than measuring $I_{Fe K\alpha}$ from the predicted spectrum, we use the emergent line intensities reported in the Cloudy output since it takes into account the effects of extinction on the line intensities. In the ND models, the Fe K$\alpha$ emission line comes from cold gas \citep[e.g][]{george91, weaver01, page04, yaqoob04, zhou05, jiang06, levenson06}, and in the WD models it comes from both cold gas and grains which are added to obtain $I_{Fe K\alpha}$. The cold Fe K$\alpha$ fluorescence line defined by Cloudy is actually emitted between 6.4 to 6.424 keV for Fe I to Fe XIV \citep{house69}; therefore, we set $\Delta E$ = 24 eV in eq. 1. As a consistency check, the EW calculation was compared with the result from Ikeda et al. (2009). Our results predict the ND \fe EW to be $\approx 1.8$ keV for $N_H$ = 10\textsuperscript{24} $\rm cm^{-2}$ and $n_H$ = 10\textsuperscript{7} $\rm cm^{-3}$ which is slightly lower than the EW $\approx$ 2 keV predicted from the Monte Carlo simulations of \cite{ikeda09}. However, given the differences in computational techniques and physical setup, this difference is adequate. In addition, we are focused on the relative changes in the EW when dust is included in the irradiated gas.

\section{Results}
\label{sec:res}

Figure 1 shows how the Fe K$\alpha$ EW depends on $N_H$ and $n_H$. The left panel of figure 1, shows that the \fe EW increases with $N_H$ for Compton thin gas (10\textsuperscript{22} $<N_H<$ 10\textsuperscript{24} cm\textsuperscript{-2}). 
\begin{figure*} 
\centerline{
\includegraphics[width=0.5\textwidth]{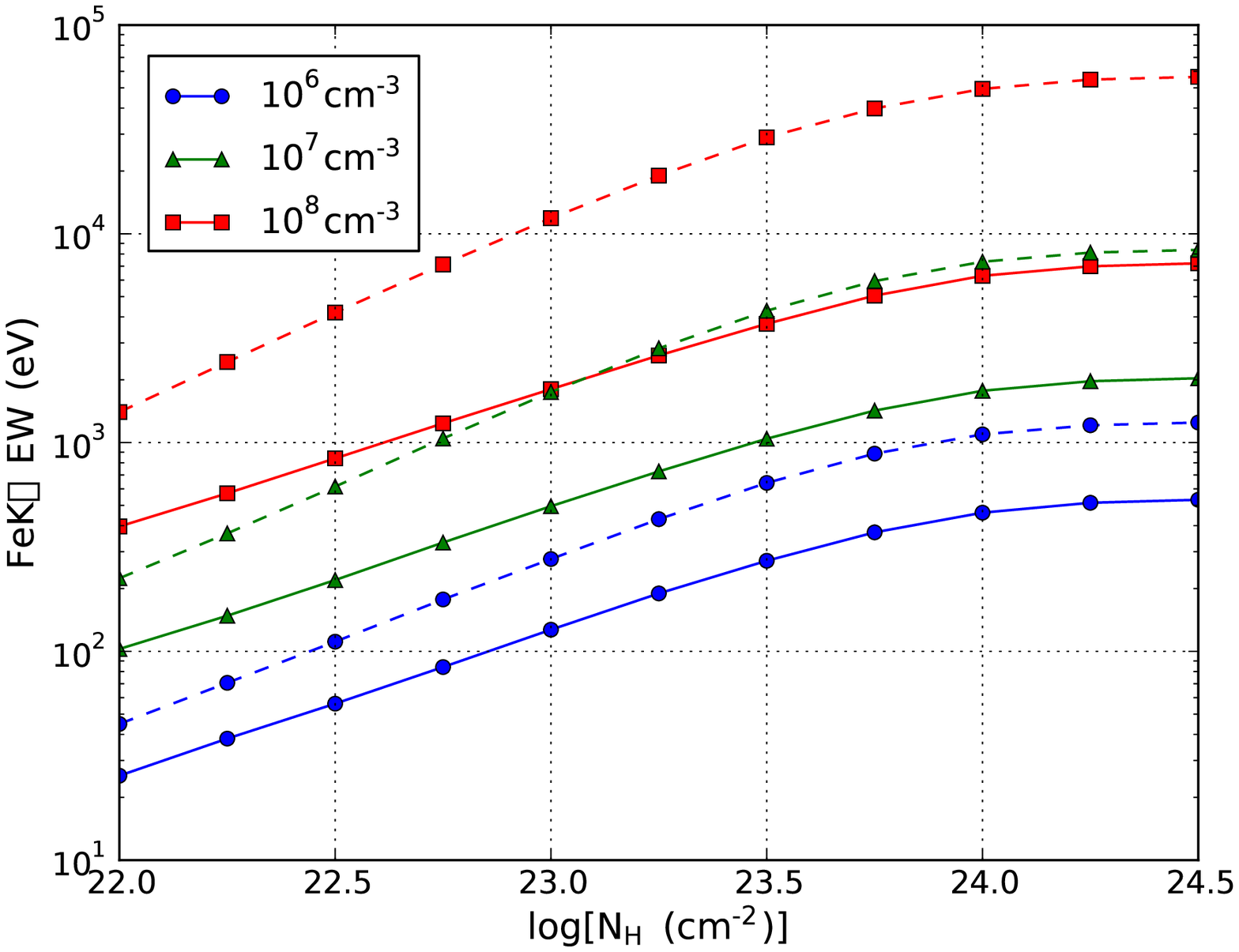}
\includegraphics[width=0.5\textwidth]{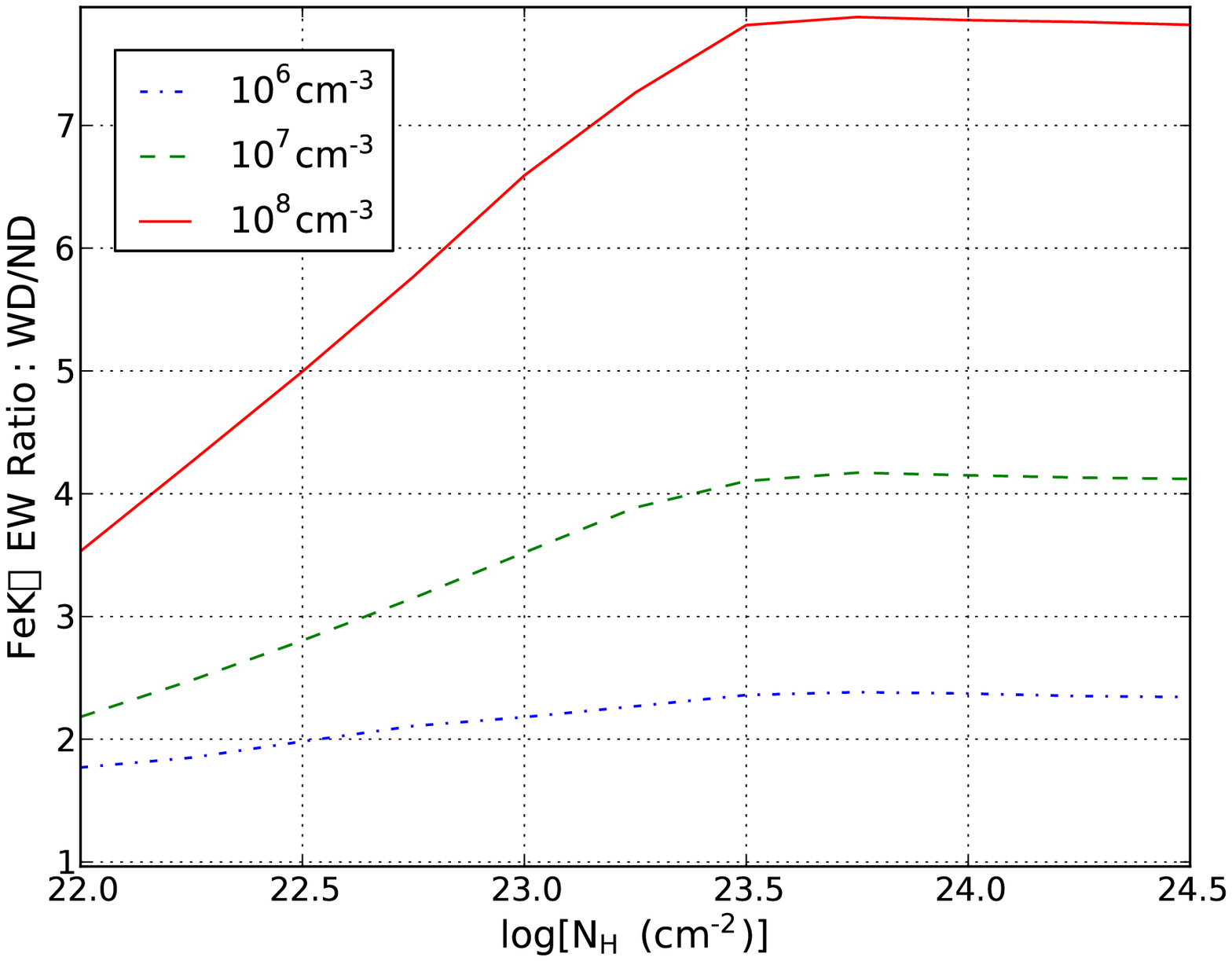}
}
\caption{$Left: $ Solid and dashed lines plot the EW versus $N_H$ for the case of ND and WD respectively. The EW increases with $N_H$ and $n_H$ in both ND and WD cases. $Right: $ The \fe EW gets significantly enhanced in the WD models with a factor that depends strongly on $n_H$.}

\label{fig:fig1}
\end{figure*}
Such a dependency is expected since the amount of illuminated gas increases with $N_H$ and this produces more \fe emission. For Compton thick gas ($N_H>$ 10\textsuperscript{24} $\rm cm^{-2}$), the amount of gas irradiated by X-rays is limited by Compton scattering and the \fe EW is fairly invariant with $N_H$. In addition, the EW of Fe K$\alpha$ increases with $n_H$ in both the ND and WD models due to the rise in total opacity with $n_H$. This reduces $I_c$ and therefore increases the EW.

Our calculation shows that the presence of dust may significantly enhance the Fe K$\alpha$ EW. The right panel of figure 1 shows that the EW of the Fe K$\alpha$ line is increased by a factor of ~2.5 for $n_H$ = 10\textsuperscript6 cm\textsuperscript{-3} (for $N_H$ $>$ 10\textsuperscript{24} cm\textsuperscript{-2}) due to presence of grains. For $n_H$ = 10\textsuperscript7 and 10\textsuperscript8 cm\textsuperscript{-3}, the EW is enhanced by a factor of ~4.1 and ~8, respectively. To illustrate the reason for this enhancement, Fig. 2 plots $I_c$ and $I_{FeK\alpha}$ ratios vs. $N_H$. This figure clearly shows that the change in EW between WD and ND is due to the reduction in the continuum and not because of a change in $I_{FeK\alpha}$. ($I_{FeK\alpha}$ does slightly increase due to the presence of grains, but this is largely due to the abundances in the models ND and WD not being perfectly self consistent; the Fe abundance is ~5\% higher in WD models. Therefore, it is expected to have slightly more Fe K$\alpha$ emission line in the case of WD.)

\begin{figure*}
\centerline{
\includegraphics[width=0.5\textwidth]{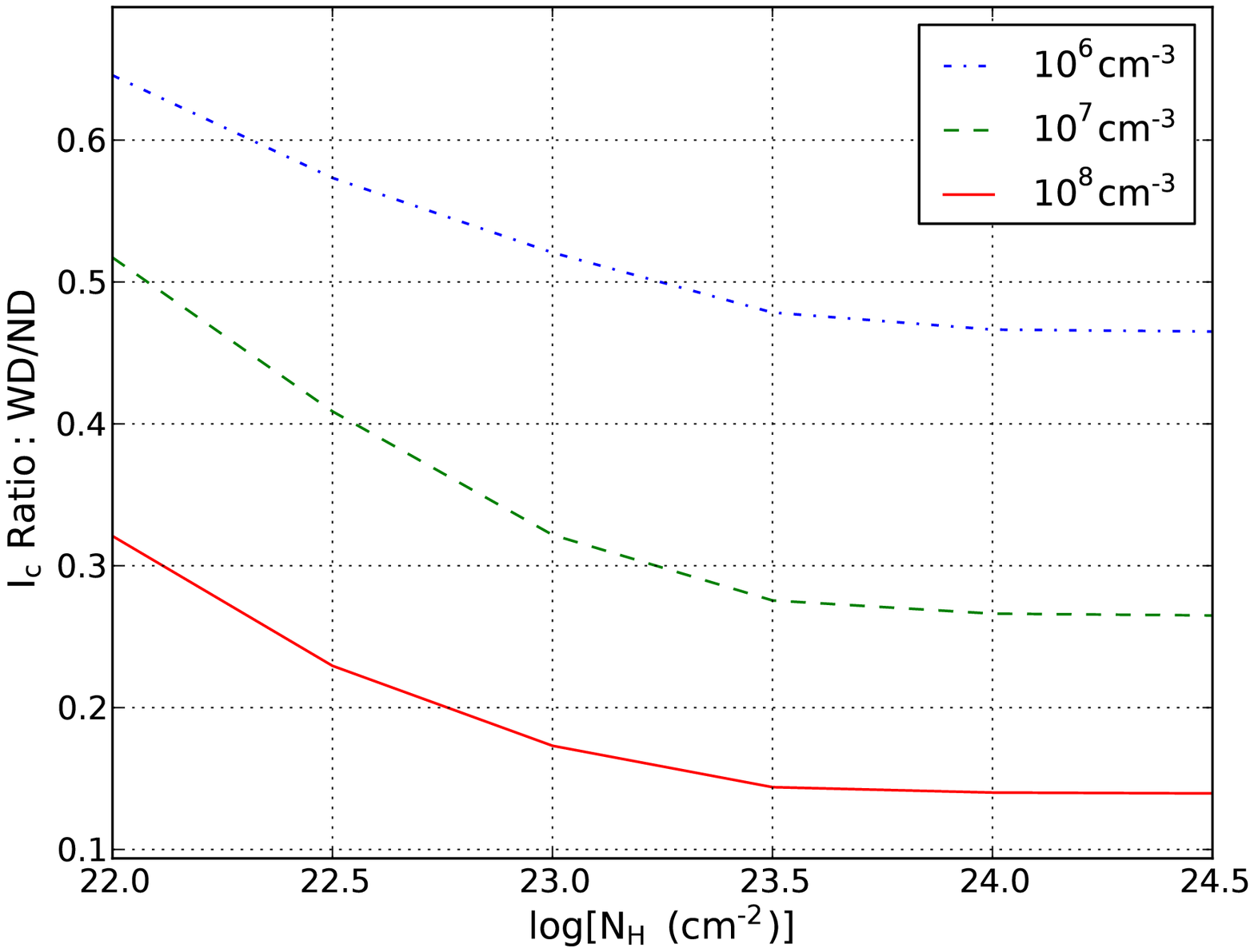}
\includegraphics[width=0.5\textwidth]{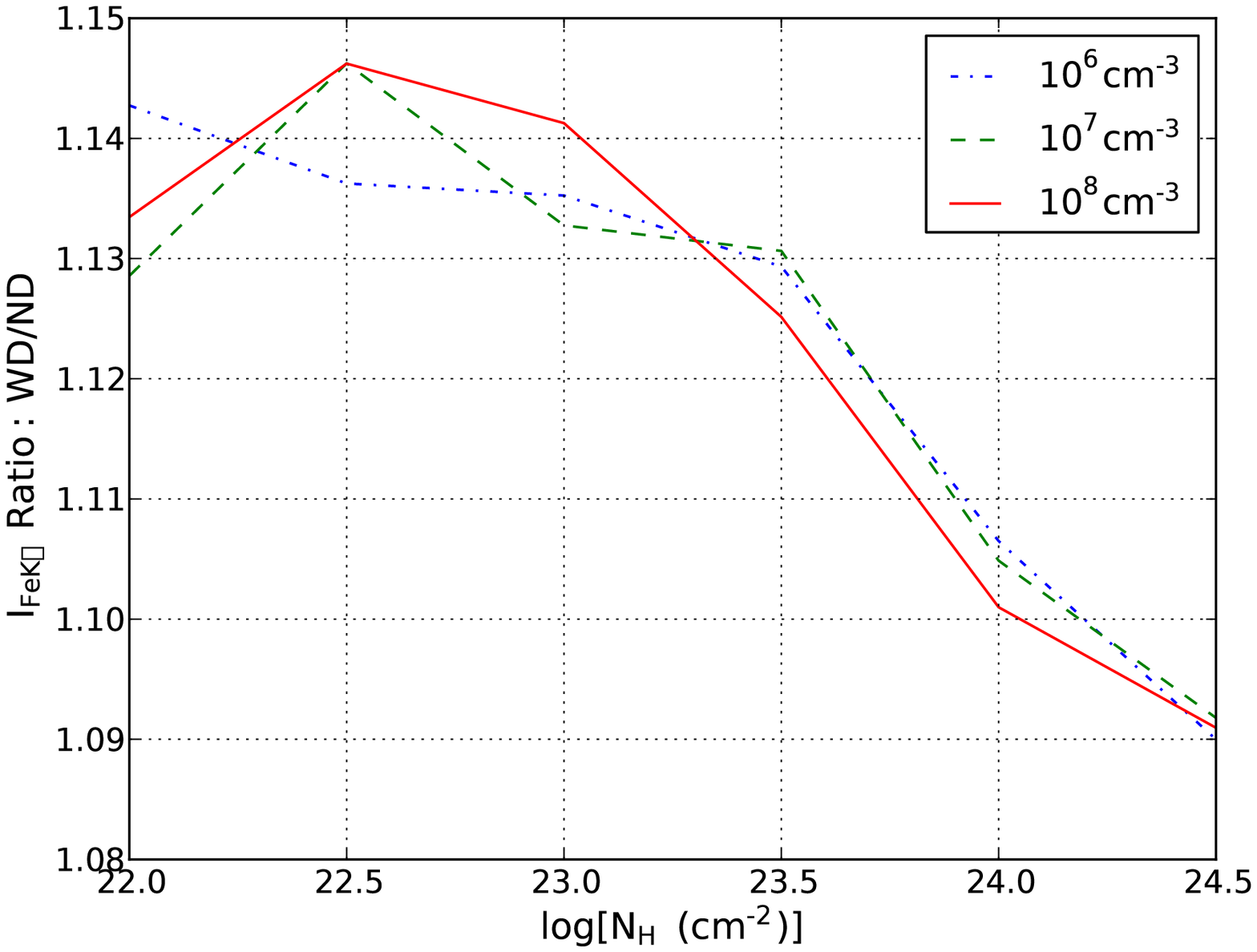}
}
\caption{$Left:$ The effect of dust on $I_c$. As $n_H$ increases, the effects of dust grows and the $I_c$ ratio significantly decreases and is therefore responsible for the increased \fe EW. $Right:$ The presence of grain only slightly increases the Fe K$\alpha$ intensity, but this has a very small effect on its EW.}
\label{fig:fig2}
\end{figure*}

Fig. 2 shows the increase in EW of Fe K$\alpha$ when reflected by dusty gas is due to a reduced continuum. This suppressed reflection continuum is due to the reduction in backscattering opacity ($\kappa_{s}$) in dusty gas. Fig. 3 shows how $\kappa_{s}$ varies inside the column of an illuminated gas as a function of depth for WD and ND models for $N_H$ = 10\textsuperscript{24} cm\textsuperscript{-2} and $n_H$ = 10\textsuperscript{7} cm\textsuperscript{-3}. At depths $>$ 10\textsuperscript{13} cm, H is no longer ionized and dust contributes significantly to the scattering opacity; therefore, $\kappa_{s}$ decreases significantly deep inside the cloud in WD models compared to the ND models. This drop in $\kappa_{s}$ due to dust is because scattering from grains is highly anisotropic and favor scattering in the forward direction by an amount depending on the size, structure and shape of the grains \citep{draine03}. Since the wavelength of X-rays is smaller than the size of most grains, there is only a weak coupling between the radiation and the grain leading to anisotropic scattering. As the Fe EW is computed using the backscattered continuum, this anisotropic scattering will reduce the backscattered intensity for WD models and thus the EW increases.

\begin{figure*}
\centerline{
\includegraphics[width=0.5\textwidth]{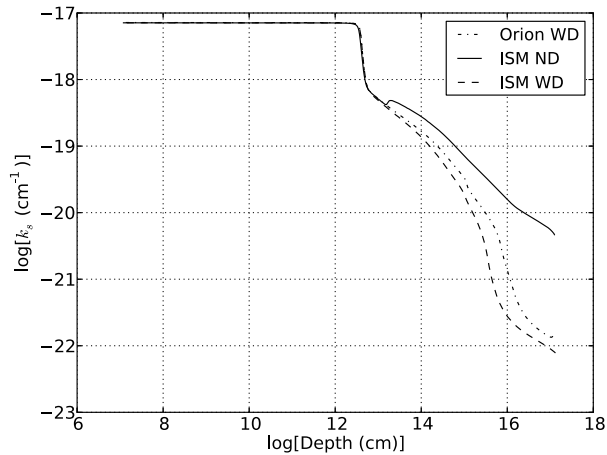}
}
\caption{The isotropic backscattering opacity in the irradiated gas as a function of depth for the $n_H$ = 10\textsuperscript{7} cm\textsuperscript{-3} and $N_H$ = 10\textsuperscript{24} cm\textsuperscript{-2} models. Deep inside the column of gas at depths $>$ 10\textsuperscript{13} cm, H recombines and the scattering is dominated by the grains which is highly anisotropic and results in lowering the backscattering opacity in the WD models. The Orion WD model shows a higher $k_s$ than the ISM WD model due to the deficiency of small grains. However, the Orion distribution still produces a lower backscattering opacity than the ND model which corresponds to an increase in the \fe EW by a factor of $\approx$ 2.6.}
\label{fig:fig3}
\end{figure*}

It has been proposed that dust in an AGN environment has fewer small grains than the typical ISM distribution \citep{maiolino01}. As $k_s$ is proportional to grain size \citep{hayakawa70}, a grain size distribution deficient in small grains will likely produce a stronger reflection continuum and thus the \fe EW will not increase as strongly as with the ISM grains. To check this effect we computed a Cloudy model ($n_H$=10\textsuperscript{7} cm\textsuperscript{-3}, $N_H$ = 10\textsuperscript{24} cm\textsuperscript{-2}) with Orion abundances that includes a grain size distribution deficient in small grains. The $k_s$ from this model is overplotted on Fig. 3 to compare against the results from the ND and WD calculations. As expected, the Orion $k_s$ is larger than the one using ISM grains, but still smaller than the model with no dust. The EW of the \fe line is enhanced by a factor of $\approx$ 4 with the ISM grains, and by a factor of $\approx$ 2.6 with the Orion grains. Hence, the anisotropic scattering behaviour of grains may still have an observable impact on the \fe EW if the smaller grains are not present.

\section{Discussion and Summary}
\label{sec:discuss}

Many groups have computed two dimensional X-ray reflection models for
AGN torii without taking into account the effects of grains
\citep[e.g.][]{ikeda09, murphy09, brightman11a}. However, it is
possible that distant X-ray reflectors arise from dusty gas
\citep{jaffe04,prieto04,prieto05,meisenheimer07,tristram07,raban09}. Moreover,
the Fe K$\alpha$ line is an important proxy to estimate the column
density of distant observing gas \citep{ghisellini94}. Therefore, we
used Cloudy to compute simple one-dimensional models of dusty gas
illuminated by an AGN in order to study the effects of dust on \fe
emission line. We found that the presence of dust may significantly
enhance the Fe K$\alpha$ EW (by factors of $\sim{5}$) in the
reflection spectrum even in non-CT gas. When grains are present in the
gas, scattering is anisotropic and there are less backscattered
photons in the reflected continuum and the overall continuum intensity
is decreased. This suggests that inferring a CT $N_H$ from the \fe EW
can be precarious. 

The increase in EW occurs when dust dominates $k_s$ and therefore
  will be most important when the gas contains predominantly neutral
  hydrogen. This limits the reflecting cloud to be relatively distant
  from the nucleus or to have a significant density. Therefore, the \fe EW
  enhacement may only be important for a certain subset of AGNs that exhibit infrared emission from AGN heated dust, a large \fe EW, and an unusually weak Compton reflection component. For example, NGC 7213 is
observed to have significant hot dust emission \citep{ruschel14}, an \fe EW $\approx 120$~eV, and a very weak Compton reflection component
\citep{bianchi09,emmanoulopoulos13}. Similary, NGC 2210 also
  shows strong dust emission \citep{honig10} and is observed to have an \fe EW of $\approx 35-200$ eV from Compton thin gas \citep{marinucci15}. Although a broad-line region origin for the \fe line is a possibility \citep{bianchi03}, our results also suggest that reprocessing from the dusty Compton thin gas in the absorber may also contribute to the \fe line.

We conclude that the anisotropic scattering behavior of grains is an important mechanism to take into account when modeling X-ray reflection from torii since it may have a significant effect on the predicted \fe EW. However, the effects of geometry are important. Since the scattering behavior of grains directly affects the continuum, the effects of grains depends on the viewing angle relative to the orientation of torus, therefore the true magnitude of the effect needs to be studied in multi-dimensional models. The EW of the emission line may be reduced, enhanced or unchanged since the continuum can be the transmitted spectrum, reflected spectrum, incident spectrum or their combination depending on the viewing angle.

\section*{Acknowledgments}
This work was supported in part by NSF awards AST 1008067 and
1333360. The authors thank P. Gandhi for comments on an earlier draft
of this work and the referee for useful comments.


\bsp 

\label{lastpage}


\begin{thebibliography}{}
\bibitem[\protect\citeauthoryear{Antonucci}{1993}]{antonucci93}
  Antonucci R. R. J., 1993, ARA\&A, 31, 473
\bibitem[\protect\citeauthoryear{Balbus}{2003}]{balbus03}
  Balbus S. A., 2003, ARA\&A, 41, 555
\bibitem[\protect\citeauthoryear{Beckmann \etal}{2009}]{beckmann09} 
  Beckmann V., Soldi S., Ricci C., 2009, A\&A, 505, 417
\bibitem[\protect\citeauthoryear{Bianchi \etal}{2009}]{bianchi09}
  Bianchi S., Guainazzi M., Matt G., Fonseca Bonilla N., Ponti G., 2009, A\&A, 495, 421
\bibitem[\protect\citeauthoryear{Bianchi \etal}{2003}]{bianchi03}
  Bianchi S., Matt G., Balestra I., Perola G. C., 2003, A\&A, 407, L21
\bibitem[\protect\citeauthoryear{Brightman \& Nandra}{2011a}]{brightman11a}
  Brightman M., Nandra K., 2011a, MNRAS, 413, 1206
\bibitem[\protect\citeauthoryear{Brusa \etal}{2005}]{brusa05}
  Brusa M., Gilli R., Comastri A., 2005, ApJ, 621, L5
\bibitem[\protect\citeauthoryear{Chaudhary \etal}{2012}]{chaudhary12}
  Chaudhary P., Brusa M., Hasinger G., Merloni A., Comastri A., Nandra K., 2012, A\&A, 537, A6
\bibitem[\protect\citeauthoryear{Corral \etal}{2011}]{corral11}
  Corral A., Della Ceca R., Caccianiga A., et al., 2011, A\&A, 530, A42
\bibitem[\protect\citeauthoryear{Corral \etal}{2008}]{corral08}
  Corral A., Page M. J., Carrera F. J., et al., 2008, A\&A, 492, 71
\bibitem[\protect\citeauthoryear{Cowie and Songaila}{1986}]{cowie86}
  Cowie L. L., Songaila A., 1986, ARA\&A, 24, 499
\bibitem[\protect\citeauthoryear{Comastri}{2004}]{comastri04}
  Comastri A., 2004, ASSL, 308, 245
\bibitem[\protect\citeauthoryear{Dadina}{2008}]{dadina08}
  Dadina M., 2008, A\&A, 485, 417
\bibitem[\protect\citeauthoryear{Draine}{2003}]{draine03}
  Draine B. T., 2003, ApJ, 598, 1026
\bibitem[\protect\citeauthoryear{Ebisawa \etal}{1996}]{ebisawa96}
  Ebisawa K., Ueda Y., Inoue H., Tanaka Y., White N.E., 1996, MNRAS, 276, 483
\bibitem[\protect\citeauthoryear{Elvis \etal}{1994}]{elvis94}
  Elvis M., Wilkes B. J., McDowell J. C., Green R. F., Bechtold J., Willner S. P., Oey M. S., Polomski E.,
  Cutri R., 1994, ApJS, 95, 1
\bibitem[\protect\citeauthoryear{Emmanoulopoulos \etal}{2013}]{emmanoulopoulos13}
  Emmanoulopolulos D., Papadakis I. E., Nicastro F., McHardy I. M., 2013, MNRAS, 429, 3439
\bibitem[\protect\citeauthoryear{Ferland \etal}{2013}]{ferland13}
  Ferland G. J., Porter R. L., Van Hoof P. A. M., Williams R. J. R., 
  Abel N.P., Lykins M. L., Gargi Shaw, Henney W. J., Stancil P. C., 2013, Revista Mexicana de Astronomia y Astrofisica, 49, 1
\bibitem[\protect\citeauthoryear{Francis}{1993}]{francis93}
  Francis P. J., 1993, ApJ, 407, 519
\bibitem[\protect\citeauthoryear{Fukazawa \etal}{2010}]{fukazawa10}
  Fukazawa Y., Hiragi K., Mizuno M., Nishino S., Hayashi K., et al., 2010, ApJ, 727, 19
\bibitem[\protect\citeauthoryear{Gandhi \etal}{2009}]{gandhi09}
  Gandhi P., Horst H., Smette A., H\"{o}nig S., Comastri A., Gilli R., Vignali C., Duschl W., 2009, A\&A, 502, 457
\bibitem[\protect\citeauthoryear{George \& Fabian}{1991}]{george91}
  George I. M., Fabian A. C., 1991, MNRAS, 249, 352
\bibitem[\protect\citeauthoryear{Ghisellini, Haardt \& Matt}{1994}]{ghisellini94}
  Ghisellini G., Haardt F., Matt G., 1994, MNRAS, 267, 743
\bibitem[\protect\citeauthoryear{Grevesse \etal}{2010}]{grevesse10}
  Grevesse N., Asplund M., Sauval A. J., Scott P., 2010, Ap\&SS, 328, 179
\bibitem[\protect\citeauthoryear{Guainazzi \etal}{2005}]{guainazzi05}
  Guainazzi M., Matt G., Perola G. C., 2005, A\&A, 444, 119
\bibitem[\protect\citeauthoryear{Hasinger}{2008}]{hasinger08}
  Hasinger G., 2008, A\&A, 490, 905
\bibitem[\protect\citeauthoryear{Hayakawa}{1970}]{hayakawa70}
  Hayakawa S., 1970, Progress of Theoretical Physics, 43, 1224
\bibitem[\protect\citeauthoryear{H\"{o}nig \etal}{2010}]{honig10}
  H{\"o}nig S.~F., Kishimoto M., Gandhi P., Smette A., Asmus D., Duschl W., Polletta M., Weigelt G., 2010, A\&A, 515, A23
\bibitem[\protect\citeauthoryear{House}{1969}]{house69}
  House L. L., 1969, ApJS, 18, 21
\bibitem[\protect\citeauthoryear{Ikeda \etal}{2009}]{ikeda09}
  Ikeda S., Awaki H., Terashima Y., 2009, ApJ, 692, 608
\bibitem[\protect\citeauthoryear{Iwasawa \etal}{2012}]{iwasawa12}
  Iwasawa K., Mainieri V., Brusa M., et al., 2012, A\&A, 537, A86
\bibitem[\protect\citeauthoryear{Jaffe \etal}{2004}]{jaffe04}
  Jaffe W., Meisenheimer K., R\"{o}ttgering, H. J. A., 2004, Nature, 429, 47
\bibitem[\protect\citeauthoryear{Jenkins}{1987}]{jenkins87}
  Jenkins E. B., 1987, Astrophysics and Space Science Library, 134, 533
\bibitem[\protect\citeauthoryear{Jiang \etal}{2006}]{jiang06}
  Jiang P., Wang J. X., Wang T. G., 2006, ApJ, 644, 725
\bibitem[\protect\citeauthoryear{Kaspi}{2001}]{kaspi01}
  Kaspi S., 2001, ApJ, 554, 216
\bibitem[\protect\citeauthoryear{Levenson \etal}{2006}]{levenson06}
  Levenson N. A., Heckman T. M., Krolik J. H., Weaver K. A., Zycki P. T., 2006, ApJ, 648, 111 
\bibitem[\protect\citeauthoryear{Maiolino}{2001}]{maiolino01}
  Maiolino, R. 2001, X-ray Astronomy: Stellar Endpoints, AGN, and the Diffuse X-ray Background, 599, 199
\bibitem[\protect\citeauthoryear{Marinucci \etal}{2015}]{marinucci15}
  Marinucci A., Matt G., Bianchi S., Lu T. N., Arevalo P., Balokovic M., Ballantyne D., Bauer F. E., et al., 2015, MNRAS, 447, 160
\bibitem[\protect\citeauthoryear{Mathis \etal}{1977}]{mathis77}
  Mathis J. S., Rumpl W., Nordsieck K. H., 1977, ApJ, 217, 425
\bibitem[\protect\citeauthoryear{Matt \etal}{2003}]{matt03}
  Matt G., Guainazzi M., Maiolino R., 2003, MNRAS, 342, 422
\bibitem[\protect\citeauthoryear{Meisenheimer \etal}{2007}]{meisenheimer07}
  Meisenheimer K., Tristram K. R. W., Jaffe W., 2007, A\&A, 471, 453
\bibitem[\protect\citeauthoryear{Meyer \etal}{1998}]{meyer98}
  Meyer D. M., Jura M., Cardelli J. A., 1998, ApJ, 493, 222
\bibitem[\protect\citeauthoryear{Murphy \& Yaqoob}{2009}]{murphy09}
  Murphy K. D., Yaqoob T., 2009, MNRAS, 397, 1549
\bibitem[\protect\citeauthoryear{Nandra \& Pounds}{1994}]{nandra94}
  Nandra K., Pounds K. A., 1994, MNRAS, 268, 405
\bibitem[\protect\citeauthoryear{Page \etal}{2004}]{page04}
  Page K. L., O’Brien P. T., Reeves J. N., Turner M. J. L., 2004, MNRAS, 347, 316
\bibitem[\protect\citeauthoryear{Prieto 2005 \etal}{2005}]{prieto05}
  Prieto, M. A., Maciejewski, W., Reunanen J., 2005, AJ, 130, 1472.
\bibitem[\protect\citeauthoryear{Prieto 2004 \etal}{2004}]{prieto04}
  Prieto M. A., Meisenhimer K., Marco O., 2004, ApJ, 614, 135
\bibitem[\protect\citeauthoryear{Raban \etal}{2009}]{raban09}
  Raban D., Jaffe E., R\"{o}ttegering H., Meisenheimer K. Tristram K. R. W., 2009, MNRAS, 394, 1325
\bibitem[\protect\citeauthoryear{Ricci \etal}{2013}]{ricci13}
  Ricci C., Paltani S., Awaki H., Petrucci P.-O., Ueda Y., Brightman M.,2013, A\&A, 553, A29
\bibitem[\protect\citeauthoryear{Ruschel-Dutra \etal}{2014}]{ruschel14}
  Ruschel-Dutra D., Pastoriza M., Riffel R., Sales D., Winge C., 2014, MRAS, 438, 3434
\bibitem[\protect\citeauthoryear{Savage and Sambach \etal}{1996}]{savage96}
  Savage B. D., Sembach K. R., 1996, AR\&A, 34, 279
\bibitem[\protect\citeauthoryear{Shu \etal}{2010}]{shu10}
  Shu X. W., Yaqoob T., Wang J. X., 2010, ApJS, 187, 581
\bibitem[\protect\citeauthoryear{Snow \etal}{2007}]{snow07}
  Snow T. P., Dodgen S. L., 1980, ApJ, 237, 708
\bibitem[\protect\citeauthoryear{Tristram \etal}{2007}]{tristram07}
  Tristram K. R. W., Meisenheimer K., Jaffe W., 2007, A\&A, 474, 837
\bibitem[\protect\citeauthoryear{Urry \& Padovani}{1995}]{urry95}
  Urry C. M., Padovani P., 1995, PASP, 107, 803
\bibitem[\protect\citeauthoryear{van Hoof \etal}{2004}]{vanhoof04}
  Van Hoof P. A. M., Weingartner J. C., Martin P. G., Volk K., Ferland G. J., 2004, MNRAS, 350, 1330
\bibitem[\protect\citeauthoryear{Weaver, Gelbord, \& Yaqoob}{2001}]{weaver01}
  Weaver K. A., Gelbord J., Yaqoob T., 2001, ApJ, 550, 261
\bibitem[\protect\citeauthoryear{Weingartner and Draine}{2001}]{draine01b}
  Weingartner J. C., Draine,2001 B. T., ApJ, 548, 296
\bibitem[\protect\citeauthoryear{Weingartner \etal}{2006}]{weingartner06}
  Weingartner J. C., Draine B. T., Barr D. K., 2006, ApJ, 645, 1188
\bibitem[\protect\citeauthoryear{Yaqoob \etal}{2001}]{yaqoob01}
  Yaqoob T., George I. M., Nandra K., Turner T. J., Serlemitsos P. J., Mushotzky R. F., 2001, ApJ, 546, 759
\bibitem[\protect\citeauthoryear{Yaqoob \& Padmanabhan}{2004}]{yaqoob04}
  Yaqoob T., Padmanabhan U. 2004, ApJ, 604, 63
\bibitem[\protect\citeauthoryear{Zamorani \etal}{1981}]{zamorani81}
  Zamorani G., Henry J. P., Maccacaro T., Tananbaum H., Soltan A., Avni Y., Liebert J., Stocke J., Strittmatter
  P. A., Weymann R. J., Smith M. G., Condon J. J., 1981, ApJ, 245, 357
\bibitem[\protect\citeauthoryear{Zhou \& Wang}{2005}]{zhou05}
  Zhou X. L., Wang J. M., 2005, ApJ, 618, L83







\end{thebibliography}
\end{document}